\documentclass{ws-procs975x65}
\begin{document}

\title{\uppercase{Perturbations of Kantowski-Sachs models}}
\author{\uppercase{Michael Bradley} $^*$ and \uppercase{Mats Forsberg}}

\address{Department of Physics, Ume\aa\ University, Sweden
\\
$^*$E-mail: michael.bradley@physics.umu.se
}

\author{\uppercase{Zolt\'{a}n Keresztes} and \uppercase{L\'{a}szl\'{o} \'{A}. Gergely}}

\address{Departments of Theoretical Physics
and %Department of 
Experimental Physics, University of Szeged, 
%Szeged, 
Hungary}

\author{\uppercase{Peter K.S. Dunsby}}

\address{Department of Mathematics and Applied Mathematics, University of Cape
Town, South Africa }

\begin{abstract}
Perturbations of Kantowski-Sachs models with a positive cosmological
constant are considered in a harmonic decomposition, in the framework of
gauge invariant 1+3 and 1+1+2 covariant splits of spacetime. Scalar, vector
and tensor modes are allowed, however they remain vorticity-free and of
perfect fluid type. The dynamics is encompassed in six evolution equations
for six harmonic coefficients.
\end{abstract}

%with a cosmological constant}}
%FOR PROCEEDINGS CONTRIBUTORS:\\ USING WORLD SCIENTIFIC'S WS-PROCS975X65\\ %DOCUMENT CLASS WITH \LaTeX2e}

%\keywords{Style file; \LaTeX; Proceedings; World Scientific Publishing.}

\bodymatter

\section{Introduction}

We investigate perturbations of Kantowski-Sachs models with a positive
cosmological constant. Although a simple argument by B\"{o}rner and Ehlers 
\cite{BornerEhlers} renders isotropic bouncing universes incompatible with
observations, the argument does not hold for Kantowski-Sachs models \cite%
{KASAscalar}, in some of which the universe exhibits an anisotropic bounce
from a contracting into an expanding phase. Hence the evolution and
propagation of perturbations in these models, also their possible effects on
observables, like the Sachs-Wolfe effect \cite{SachsWolfe} are of interest.

The 1+3 and 1+1+2 covariant splits of the space-time \cite{EllisBruni,
EllisvanElst,ClarksonBarrett, Clarkson} prove particularly suitable for
perturbation theory, as they employ variables vanishing on the background,
hence with gauge invariant perturbations \cite{StewartWalker}. For
simplicity we consider only vorticity-free perturbations of perfect fluid
type, but allow for scalar, vector and tensor modes. We further expand the
evolution equations for the perturbative variables in terms of harmonics.

\section{The 1+3 and 1+1+2 covariant formalisms}

The existence of a preferred timelike vector $u^{a}$ naturally leads to a
covariant 1+3 split of spacetimes \cite{EllisBruni, EllisvanElst}. The
induced metric $h_{a}^{b}=g_{a}^{b}+u_{a}u^{b}$ also acts as a projection
operator onto the perpendicular 3-space. All vectors and tensors can be
covariantly decomposed into "spatial" and "timelike" parts. Covariant time
derivative and projected spatial derivative can be defined as 
\[
\dot{\psi}_{a..b}\equiv u^{c}\nabla _{c}\psi _{a...b}\quad \hbox{and}\quad
D_{c}\psi _{a...b}\equiv h_{c}^{f}h_{a}^{d}...h_{b}^{e}\nabla _{f}\psi
_{d...e} 
\]%
respectively. The covariant derivative of the 4-velocity $u^{a}$ is
decomposed as 
\begin{equation}
\nabla _{a}u_{b}=-u_{a}A_{b}+D_{a}u_{b}=-u_{a}A_{b}+\frac{1}{3}\theta
h_{ab}+\omega _{ab}+\sigma _{ab}
\end{equation}%
in terms of the kinematic quantities of $u^{a}$, the acceleration $%
A_{a}\equiv u^{b}\nabla _{b}u_{a}$, expansion $\theta \equiv D_{a}u^{a}$,
vorticity $\omega _{ab}\equiv D_{[a}u_{b]}$ and shear $\sigma _{ab}\equiv
D_{<a}u_{b>}$ respectively. The kinematic quantities, supplemented by the
Ricci tensor (expressed via the Einstein equations in terms of the energy
density $\mu $ and pressure $p$ of a perfect fluid) and by the electric and
magnetic parts of the Weyl tensor ($E_{ab}\equiv C_{acbd}u^{c}u^{d}$ and $%
H_{ab}\equiv \frac{1}{2}\eta _{ade}C^{de}\!\!_{bc}u^{c}$)\ are the dependent
variables. The Ricci and Bianchi identities lead to both evolution equations
in the $u^{a}$ direction and to constraints.

A formalism for a further split (1+2) with respect to a spatial vector $%
n^{a} $ (with $u^{a}n_{a}=0$) is also known \cite{ClarksonBarrett, Clarkson}%
. The projections perpendicular to $n^{a}$ arise via the induced metric $%
N_{a}^{b}=h_{a}^{b}-n_{a}n^{b}$ of the 2-space. Vectors and tensors may be
decomposed again into scalars along $n^{a}$, perpendicular two-vectors and
symmetric, trace-free two-tensors as $A^{a}=\mathcal{A}n^{a}+\mathcal{A}^{a}$
, $\omega ^{a}=\Omega n^{a}+\Omega ^{a}$, $\sigma _{ab}=\Sigma (n^{a}n^{b}-%
\frac{1}{2}N_{ab})+2\Sigma _{(a}n_{b)}+\Sigma _{ab}$ and similarly for $%
E_{ab}$ and $H_{ab}$ in terms of $\mathcal{E}$, $\mathcal{E}_{a}$, $\mathcal{%
E}_{ab}$ and $\mathcal{H}$, $\mathcal{H}_{a}$, $\mathcal{H}_{ab}$
respectively. The derivatives along and perpendicular to $n^{a}$ are%
\[
\hat{\psi}_{a...b}\equiv n^{c}D_{c}\psi
_{a...b}=n^{c}h_{c}^{f}h_{a}^{d}...h_{b}^{e}\nabla _{f}\psi _{d...e}\;\;%
\hbox{and}\;\;\delta _{c}\psi _{a...b}\equiv
N_{c}^{f}N_{a}^{d}...N_{b}^{e}D_{f}\psi _{d...e} 
\]%
respectively. Similarly to $\nabla _{a}u_{b}$, the derivatives $D_{a}n_{b}$
and $\dot{n}_{a}$ can be decomposed in terms of `kinematical' quantities
related to $n^{a}$ as 
\begin{equation}
D_{a}n_{b}=n_{a}a_{b}+\frac{1}{2}\phi N_{ab}+\xi \epsilon _{ab}+\zeta
_{ab}\quad \hbox{and}\quad \dot{n}_{a}=\mathcal{A}u_{a}+\alpha _{a}
\end{equation}%
where $a_{a}\equiv \hat{n}_{a}$, $\phi \equiv \delta _{a}n^{a}$, $\xi \equiv 
\frac{1}{2}\epsilon ^{cabd}\delta _{a}n_{b}u_{c}n_{d}$, $\zeta _{ab}\equiv
\delta _{\{a}n_{b\}}$, $\mathcal{A}\equiv n^{a}A_{a}$, $\alpha _{a}\equiv
N_{a}^{b}\dot{n}_{b}$.

As summary, the Ricci and Bianchi identities can be rewritten as evolution
and propagation equations in the $u^{a}$ and $n^{a}$ directions plus
constraints.

\section{Perturbations of Kantowski-Sachs}

For the unperturbed space-time we take Locally Rotationally Symmetric (LRS)
Kantowski-Sachs cosmologies \cite{KantowskiSachs} 
\begin{equation}
ds^{2}=-dt^{2}+a_{1}^{2}(t)dz^{2}+a_{2}^{2}(t)\left( d\vartheta ^{2}+\sin
^{2}\theta d\varphi ^{2}\right) \,
\end{equation}%
with cosmological constant $\Lambda >0$ and a perfect fluid with barytropic
equation of state $p=p(\mu )$. These space-times are completely determined
by the shear $\Sigma $, energy density $\mu $, expansion $\theta $. The
electric part of the Weyl tensor $\mathcal{E}$ is given algebraically in
terms of the others. %evolve as
%\begin{eqnarray}\label{eq01}
%\dot \Sigma =-\frac{1}{2}\Sigma^2-\frac{2}{3}\Sigma\theta-{\cal E},
% \;\;
%\dot \mu=-\theta (\mu + p), \;\;
%\dot \theta =(\Lambda-\frac{1}{2}\mu-\frac{3}{2}p) -\frac{1}{3}\theta^2-\frac{3}{2}\Sigma^2 
%\end{eqnarray}
%where the electric part of the Weyl tensor is 
%${\cal E}=- \frac{2}{3}\mu- \frac{2}{3}\Lambda-\Sigma^2+\frac{2}{9}\theta^2+\frac{1}{3}\Sigma\theta$.

The perturbative analysis proceeds smoothly in terms of variables vanishing
on the background, hence being gauge invariant \cite{StewartWalker}. The
spatial variations in $\theta ,\Sigma ,\mathcal{E},\mu $ are characterized
by their gradients 
\begin{equation}
W_{a}\equiv \delta _{a}\theta ,\quad V_{a}\equiv \delta _{a}\Sigma ,\quad
X_{a}\equiv \delta _{a}\mathcal{E},\quad \mu _{a}\equiv \delta _{a}\mu \,,
\end{equation}%
vanishing on the background (the derivatives $\hat{\theta}\equiv
n^{a}D_{a}\theta $ etc. can be rewritten in terms of $\delta _{a}$%
-derivatives due to the commutation relations holding in the case of no
vorticity). Other variables vanishing on the background are $a_{a},\phi ,\xi
,\zeta _{ab},\alpha _{a},\mathcal{A},\mathcal{A}_{a}$, $\Sigma _{a},\Sigma
_{ab},\mathcal{E}_{a},\mathcal{E}_{ab},\mathcal{H},\mathcal{H}_{a},\mathcal{H%
}_{ab}$, where $a_{a}$ can be put to zero by choice of frame.

Any scalar, vector and tensor variable is expanded in harmonics according to 
\begin{eqnarray}
\Psi &=&\sum\limits_{k_{\parallel },k_{\perp }}\Psi _{k_{\parallel }k_{\perp
}}P_{k_{\parallel }}Q_{k_{\perp }}\,,\quad \Psi
_{a}=\sum\limits_{k_{\parallel },k_{\perp }}P_{k_{\parallel }}\left( \Psi
_{k_{\parallel }k_{\perp }}^{V}Q_{a}^{k_{\perp }}+\overline{\Psi }%
_{k_{\parallel }k_{\perp }}^{V}\overline{Q}_{a}^{k_{\perp }}\right) \,, 
\nonumber \\
\Psi _{ab} &=&\sum\limits_{k_{\parallel },k_{\perp }}P_{k_{\parallel }}\
\left( \Psi _{k_{\parallel },k_{\perp }}^{T}Q_{ab}^{k_{\perp }}+\overline{%
\Psi }_{k_{\parallel },k_{\perp }}^{T}\overline{Q}_{ab}^{k_{\perp }}\right)
\end{eqnarray}%
where $Q_{k_{\perp }}$, $Q_{a}^{k_{\perp }}$, $\overline{Q}_{a}^{k_{\perp }}$%
, $Q_{ab}^{k_{\perp }}$ and $\overline{Q}_{ab}^{k_{\perp }}$ are harmonics
on the 2-spheres of constant $z$ and $P_{k_{\parallel }}$ are the
corresponding expansion functions in the $z$-direction.

All first order quantities can be given in terms of six coefficients. Four
of them, $\mu _{k_{\parallel },k_{\perp }}^{V}$, $\Sigma _{k_{\parallel
},k_{\perp }}^{T}$, $\mathcal{E}_{k_{\parallel },k_{\perp }}^{T}$ and $%
\overline{\mathcal{H}}_{k_{\parallel },k_{\perp }}^{T}$ form a closed first
order system of evolution equations coupled to the density gradient, in
agreement with earlier results for scalar perturbations \cite{KASAscalar}.
The two other $\overline{\mathcal{E}}_{k_{\parallel },k_{\perp }}^{T}$ and $%
\mathcal{H}_{k_{\parallel },k_{\perp }}^{T}$ form a closed system for free
waves.

These sets of equations seem suitable to study the propagation of
gravitational waves, and the coupling between scalar and tensor
perturbations. Furthermore, the null geodesic condition for photons leads to
equations giving the redshift in different directions in terms of the 1+1+2
quantities. Hence the Sachs-Wolfe effect and the corresponding variations in
the CMB temperature can be calculated.

\section*{Acknowledgements}

Zolt\'{a}n Keresztes was supported by OTKA grant no. 100216

\end{document}